\documentclass[twocolumn,aps,prl,amsmath,showpacs]{revtex4}

\usepackage{latexsym}
\usepackage{graphicx}
\usepackage{amsfonts}

\usepackage{amssymb}                                                          

\begin{document}

\title{Fluctuation relations without microscopic time reversality:
Generalized Green-Kubo relation and integral fluctuation theorem
for uniformly sheared granular systems}

\author{Song-Ho Chong$^{1}$, Michio Otsuki$^{2}$ and Hisao Hayakawa$^{3}$}
\affiliation{
$^{1}$Institute for Molecular Science,
      Okazaki 444-8585, Japan \\
$^{2}$Department of Physics and Mathematics, Aoyama Gakuin University,
      Kanagawa 229-8558, Japan\\
$^{3}$Yukawa Institute for Theoretical Physics, Kyoto University, Kyoto 606-8502, Japan}
\date{\today}

\begin{abstract}

We derive the generalized Green-Kubo relation and an integral form of the fluctuation theorem
that apply to uniformly sheared granular systems in which microscopic
time-reversal symmetry is broken.
The former relation provides an exact representation of nonequilibrium steady-state 
properties as the integral of the time-correlation function describing transient dynamics from 
an initial quiescent towards a final sheared steady state.
We also investigate implications of the integral fluctuation theorem
on the approach towards the steady state and on the possible form of the
steady-state distribution function in terms of the excess thermodynamic function. 

\end{abstract}


\maketitle

Developing statistical mechanics for 
nonequilibrium steady states is one of the most challenging 
problems in physics. 
Indeed, quite a few studies have been devoted to such a 
development~\cite{Zubarev74,McLennan88,Evans08,Kawasaki73,Morriss87,FT-papers,Jarzynski97,Evans02,Seifert05,Komatsu-all},
and among the most remarkable outcomes have been the
generalized Green-Kubo relation~\cite{Morriss87}
and various forms of fluctuation theorems~\cite{FT-papers,Jarzynski97,Evans02,Seifert05}.
A suggestive representation of the steady-state distribution
in terms of the excess entropy production
has been recently proposed~\cite{Komatsu-all}, and its connection to the
steady-state thermodynamics has been discussed~\cite{Komatsu08b}. 
However, most of these studies rest on the use of microscopic
time reversality or the local detailed balance, and 
they cannot be applied literally to 
macroscopic dissipative systems like granular fluids~\cite{Puglisi05} 
despite manifest similarities~\cite{Feitosa04}.
In this Letter, we explore to what extent 
those outcomes hold also for this important class of systems
in which microscopic time reversality is broken. 
It is demonstrated that the generalized Green-Kubo relation,
an integral form of the fluctuation theorem, and the steady-state 
distribution in terms of the excess thermodynamic function can be derived
without resorting to microscopic time reversality.
This will be exemplified for uniformly sheared granular systems.
Such a formulation will also be useful in studies of
jammed glassy materials~\cite{jamming-all}.

We shall consider a system of $N$ dissipative soft-sphere particles
of mass $m$ in a volume $V$
subjected to stationary shearing characterized by the
shear-rate tensor $\kappa_{\lambda \mu} = \dot{\gamma} \delta_{\lambda x} \delta_{\mu y}$
with the shear rate $\dot{\gamma}$. 
It is assumed that the applied shear induces a homogeneous 
streaming-velocity profile 
$\mbox{\boldmath $\kappa$}\cdot {\bf r}$
at position ${\bf r}$.
Equations of motion describing such a system
are the SLLOD equations~\cite{Evans08}, 
\begin{subequations}
\label{eq:SLLOD}
\begin{eqnarray}
\dot{\bf r}_{i} &=& \frac{{\bf p}_{i}}{m} +
\mbox{\boldmath $\kappa$} \cdot {\bf r}_{i},
\label{eq:SLLOD-a}
\\
\dot{\bf p}_{i} &=& \sum_{j \ne i} {\bf F}_{ij} -
\mbox{\boldmath $\kappa$} \cdot {\bf p}_{i}. 
\label{eq:SLLOD-b}
\end{eqnarray}
\end{subequations}
Here ${\bf r}_{i}$ and ${\bf p}_{i}$ refer to the position and peculiar
momentum of particle $i$. 
We assume pairwise-additive ``smooth'' contact forces acting only
on the normal direction.
The simplest realistic model for such a 
force ${\bf F}_{ij}$ that particle $j$ exerts on
particle $i$ is the frictional contact model~\cite{Dufty-all},
\begin{equation}
{\bf F}_{ij} =
\hat{\bf r}_{ij}
\Theta(d-r_{ij}) [ f(d-r_{ij}) - \gamma(d-r_{ij}) ({\bf g}_{ij} \cdot \hat{\bf r}_{ij})].
\label{eq:Fij}
\end{equation}
Here $d$ denotes the particle diameter;
$\Theta(x)$ is the Heaviside step function;
${\bf r}_{ij} \equiv {\bf r}_{i} - {\bf r}_{j}$ and 
${\bf g}_{ij} \equiv (\dot{\bf r}_{i} - \dot{\bf r}_{j})/m$;
and $\hat{\bf r}_{ij} \equiv {\bf r}_{ij} / r_{ij}$
with $r_{ij} \equiv | {\bf r}_{ij} |$.
The first term in Eq.~(\ref{eq:Fij})
describes a conservative force representing
the elastic repulsion: typical functional forms are 
$f(x) \propto x$ (linear model) and
$f(x) \propto x^{3/2}$ (Hertzian contact model).
The second term refers to a nonconservative, dissipative force
due to inelastic collisions.
It is proportional to the relative velocity of colliding particles,
and breaks the time-reversal symmetry of Eq.~(\ref{eq:SLLOD-b}).
The amount of energy dissipation
is characterized by the viscous function $\gamma(x)$ which is 
typically assumed to be constant or modeled as 
$\gamma(x) \propto x^{1/2}$.
The internal energy of the system is 
\begin{equation}
H_{0} = 
\sum_{i} \frac{{\bf p}_{i}^{2}}{2m} + 
\frac{1}{2}
\sum_{i} \sum_{j \ne i} V(r_{ij}),
\end{equation}
where the potential energy function $V(r_{ij})$ satisfies 
$\partial V(r_{ij}) / \partial r_{ij} =
- \Theta(d-r_{ij}) f(d-r_{ij})$.
The rate of the internal energy change is computed by using 
Eqs.~(\ref{eq:SLLOD}) as
\begin{equation}
\dot{H}_{0} = - \dot{\gamma} \sigma_{xy} - 2 {\cal R}.
\label{eq:dot-H0}
\end{equation}
Here we have used the specific form $\kappa_{\lambda \mu} = 
\dot{\gamma} \delta_{\lambda x} \delta_{\mu y}$
and the symmetry
$\sigma_{\lambda \mu} = \sigma_{\mu \lambda}$ of 
the stress tensor 
\begin{equation}
\sigma_{\lambda \mu} =
\sum_{i} 
\Bigl[ \, 
  \frac{p_{i, \lambda} p_{i, \mu}}{m} + r_{i,\lambda} \sum_{j \ne i} F_{ij, \mu} \,
\Bigr] .
\label{eq:sigma-def}
\end{equation}
${\cal R}$ is Rayleigh's dissipation function~\cite{Goldstein80}
\begin{equation}
{\cal R} = \frac{1}{4} \sum_{i} \sum_{j \ne i}
\Theta(d-r_{ij}) \gamma(d-r_{ij}) ({\bf g}_{ij} \cdot \hat{\bf r}_{ij})^{2}.
\label{eq:R-def}
\end{equation}

The time evolution of phase variables, i.e., functions of 
the phase-space point
${\bf \Gamma} = ({\bf r}^{N}, {\bf p}^{N})$, 
is determined by the Liouville equation~\cite{Evans08}
\begin{equation}
\frac{d}{dt} A({\bf \Gamma}) =
\dot{\bf \Gamma} \cdot 
\frac{\partial}{\partial {\bf \Gamma}}
A({\bf \Gamma}) \equiv
i {\cal L} A({\bf \Gamma}).
\label{eq:Lp}
\end{equation}
The explicit expression for the Liouville operator $i {\cal L}$
can easily be found 
from Eqs.~(\ref{eq:SLLOD}).
A formal solution to Eq.~(\ref{eq:Lp}) reads
$A({\bf \Gamma}(t)) = \exp(i {\cal L} t) A({\bf \Gamma})$.
(Hereafter, the absence of the argument $t$ implies that
associated quantities are evaluated at $t=0$,
and the dependence on ${\bf \Gamma}$ shall often be dropped 
for brevity.)
On the other hand, 
the Liouville equation for the 
phase-space distribution function $f({\bf \Gamma},t)$ is given by~\cite{Evans08}
\begin{equation}
\frac{\partial f({\bf \Gamma},t)}{\partial t} =
- [ \, i {\cal L} + \Lambda({\bf \Gamma}) \, ] \, f({\bf \Gamma},t) \equiv
- i {\cal L}^{\dagger} f({\bf \Gamma},t).
\label{eq:Lf}
\end{equation}
Here $\Lambda({\bf \Gamma}) \equiv
(\partial / \partial {\bf \Gamma}) \cdot
\dot{\bf \Gamma}$ is the phase-space compression factor.
For the SLLOD equations (\ref{eq:SLLOD}), one obtains
\begin{equation}
\Lambda({\bf \Gamma}) =
- \frac{1}{m} \sum_{i} \sum_{j \ne i}
\Theta(d-r_{ij}) \gamma(d-r_{ij}) \le 0.
\label{eq:SLLOD-Lambda}
\end{equation}
A formal solution to Eq.~(\ref{eq:Lf}) is
$f({\bf \Gamma},t) = \exp( - i {\cal L}^{\dagger} t) \,
f({\bf \Gamma})$.
In the following, we shall often use the relations~\cite{Evans08}
\begin{eqnarray}
\int d{\bf \Gamma} \,
[ e^{i {\cal L} t} A({\bf \Gamma})] \, B({\bf \Gamma}) =
\int d{\bf \Gamma} \,
A({\bf \Gamma}) \, [ e^{- i {\cal L}^{\dagger} t} B({\bf \Gamma})],
\label{eq:unrolling}
\\
\exp( - i {\cal L}^{\dagger} t) =
\exp\Bigl[ - \int_{0}^{t} ds \, \Lambda(-s) \Bigr]
\exp( - i {\cal L} t).
\label{eq:relation-f-p-propagators} 
\end{eqnarray}

Let us consider the following 
realization of the nonequilibrium steady state.
The system is first equilibrated at the inverse temperature 
$\beta = 1/T$ (setting Boltzmann's constant unity)
by turning off the shearing and dissipative forces.
The distribution function for such a fictitious state
is given by the canonical one
\begin{equation}
f_{\rm c}({\bf \Gamma}; \beta) \equiv 
\frac{e^{- \beta H_{0}({\bf \Gamma})}}{\cal Z(\beta)}, \,\,\,
{\cal Z}(\beta) = \int d{\bf \Gamma} \,
e^{- \beta H_{0}({\bf \Gamma})}.
\label{eq:fc-def}
\end{equation}
This choice of the fictitious initial state will be justified below. 
At time $t=0$, the shearing and dissipative forces are turned on, and thereafter
the system evolves according to the SLLOD equations~(\ref{eq:SLLOD}).
The nonequilibrium distribution function for $t > 0$ is therefore given by
\begin{equation}
f({\bf \Gamma},t) = \exp( - i {\cal L}^{\dagger} t) \,
f_{\rm c}({\bf \Gamma}; \beta).
\label{eq:f-propagator-beta}
\end{equation}
The steady state is assumed to be reached for $t \to \infty$. 

One obtains from Eqs.~(\ref{eq:relation-f-p-propagators}) and (\ref{eq:f-propagator-beta})
the so-called Kawasaki representation~\cite{Evans08}
\begin{eqnarray}
f({\bf \Gamma},t) &=&
\exp\Bigl[ - \int_{0}^{t} ds \, \Lambda(-s) \Bigr]
\frac{e^{-\beta H_{0}(-t)}}{\cal Z(\beta)}
\nonumber \\
&=&
f_{\rm c}({\bf \Gamma}; \beta)
\exp\Bigl[ \int_{0}^{t} ds \, \Omega(-s) \Bigr].
\label{eq:new-neq-distribution}
\end{eqnarray}
In the second equality we have used 
$H_{0}(-t) = H_{0}(0) - \int_{0}^{t} ds \, \dot{H}_{0}(-s)$
and Eq.~(\ref{eq:dot-H0}), 
and introduced
\begin{equation}
\Omega({\bf \Gamma}) \equiv
-\beta \dot{\gamma} \sigma_{xy}({\bf \Gamma}) - 
2 \beta {\cal R}({\bf \Gamma})  - \Lambda({\bf \Gamma}).
\label{eq:Omega-def}
\end{equation}
One easily finds that $\Omega({\bf \Gamma})$ coincides with the
dissipation function introduced in Ref.~\cite{Evans02}.
From the normalization $\int d{\bf \Gamma} \, f({\bf \Gamma},t) = 1$,
one obtains a Jarzynski-type equality
$\langle e^{\int_{0}^{t} ds \, \Omega(-s)} \rangle_{\beta} = 1$~\cite{Jarzynski97},
but a more useful form of such an equality shall be derived below.
From here on, 
$\langle \cdots \rangle_{\rm \beta} \equiv 
\int d{\bf \Gamma} \, f_{\rm c}({\bf \Gamma}; \beta) \cdots$
refers to the averaging over the initial canonical distribution function.

For the  nonequilibrium average 
$\langle A(t) \rangle_{\beta}$ defined by
\begin{equation}
\langle A(t) \rangle_{\beta} \equiv
\int d{\bf \Gamma} \,
f_{\rm c}({\bf \Gamma}; \beta) \, A(t) = 
\int d{\bf \Gamma} \, A(0) \,
f({\bf \Gamma},t),
\label{eq:neq-average}
\end{equation}
in which the second equality follows from Eq.~(\ref{eq:unrolling}),
one finds using Eq.~(\ref{eq:new-neq-distribution})
\begin{equation}
\langle A(t) \rangle_{\beta} =
\langle A(0) e^{\int_{0}^{t} ds \, \Omega(-s)} \rangle_{\beta}.
\label{eq:dissipation-theorem-1}
\end{equation}
By differentiating and then integrating this equation
with respect to time, we obtain
\begin{equation}
\langle A(t) \rangle_{\beta} = \langle A(0) \rangle_{\beta} +
\int_{0}^{t} ds \, \langle A(s) \Omega(0) \rangle_{\beta}.
\label{eq:dissipation-theorem-2}
\end{equation}
The derivation of Eqs.~(\ref{eq:dissipation-theorem-1}) and (\ref{eq:dissipation-theorem-2})
in terms of the dissipation function defined in Eq.~(\ref{eq:Omega-def}) is
called the dissipation theorem~\cite{Evans08b}.
Although it is implied in Ref.~\cite{Evans08b} that the dissipation theorem
is a corollary of the transient fluctuation theorem resting on 
microscopic time reversality, 
it holds also in the absence of such a symmetry.

However, a question arises as to the utility of 
Eqs.~(\ref{eq:dissipation-theorem-1}) and (\ref{eq:dissipation-theorem-2})
since they explicitly refer to the fictitious initial equilibrium distribution.
In this regard, let us prove here two important properties concerning 
$\lim_{t \to \infty} \langle A(t) \rangle_{\beta}$
for systems that exhibit mixing~\cite{comment-mixing}.
First, it follows from Eq.~(\ref{eq:dissipation-theorem-2}) for $t \to \infty$
\begin{equation}
\frac{d}{dt} \langle A(t) \rangle_{\beta} =
\langle A(t) \Omega(0) \rangle_{\beta} \to 
\langle A(t) \rangle_{\beta} \langle \Omega(0) \rangle_{\beta} = 0,
\label{eq:mixing-1}
\end{equation}
meaning that
$\lim_{t \to \infty} \langle A(t) \rangle_{\beta}$ becomes a constant.
Here we have used 
$\langle \Omega(0) \rangle_{\beta} = 0$ which can easily be confirmed
from Eqs.~(\ref{eq:sigma-def}), (\ref{eq:R-def}), and (\ref{eq:SLLOD-Lambda}).
Thus, the steady-state average 
$\langle A \rangle_{\rm ss} \equiv \lim_{t \to \infty} \langle A(t) \rangle_{\beta}$
is well defined, and is given by setting $t \to \infty$ in Eq.~(\ref{eq:dissipation-theorem-2}) as
\begin{equation}
\langle A \rangle_{\rm ss} = \langle A(0) \rangle_{\beta} +
\int_{0}^{\infty} ds \, \langle A(s) \Omega(0) \rangle_{\beta}.
\label{eq:ss-average-TTCF}
\end{equation}
Second, since 
\begin{equation}
\frac{\partial}{\partial \beta} 
\Bigl\{ \frac{e^{-\beta H_{0}(-t)}}{\cal Z(\beta)} \Bigr\} =
[ \langle H_{0} \rangle_{\beta} - H_{0}(-t) ] \, 
\frac{e^{-\beta H_{0}(-t)}}{\cal Z(\beta)},
\label{eq:dA-dbeta-dum-03}
\end{equation}
one obtains from Eqs.~(\ref{eq:new-neq-distribution}) and (\ref{eq:neq-average})
for $t \to \infty$
\begin{eqnarray}
\frac{\partial}{\partial \beta} 
\langle A(t) \rangle_{\beta} &=&
\int d{\bf \Gamma} \, 
A(0) \, [ \langle H_{0} \rangle_{\beta} - H_{0}(-t) ] \, f({\bf \Gamma},t)
\nonumber \\
&=&
\langle A(t) \rangle_{\beta} 
\langle H_{0} \rangle_{\beta} -
\langle A(t) H_{0} \rangle_{\beta} \to 0,
\label{eq:dA-dbeta-dum-04}
\end{eqnarray}
i.e., 
$\langle A \rangle_{\rm ss} = \lim_{t \to \infty} \langle A(t) \rangle_{\beta}$
is in fact independent of the inverse temperature $\beta$
of the fictitious initial equilibrium state.
Actually, 
one can prove a stronger statement~\cite{Otsuki-proof}:
$\langle A \rangle_{\rm ss}$ determined from Eq.~(\ref{eq:ss-average-TTCF}) coincides
with the steady-state average calculated based on an arbitrary initial 
distribution function $f({\bf \Gamma})$ via
$\langle A \rangle_{\rm ss} =
\lim_{t \to \infty}
\int d{\bf \Gamma} \,
A(0) \exp(- i {\cal L}^{\dagger} t) f({\bf \Gamma})$.
Thus, $\langle A \rangle_{\rm ss}$ 
is uniquely specified by the ``thermodynamic'' parameters
$(N, V, \dot{\gamma})$ characterizing the steady state,
irrespective of the initial distribution function.
It is therefore most convenient to adopt the 
aforementioned fictitious state for which the canonical
distribution function is available.

Equation~(\ref{eq:ss-average-TTCF}) is the generalized
Green-Kubo formula relating the steady-state average to the
time-correlation function describing transient dynamics from 
an initial equilibrium towards a final steady state.
It is a natural extension of the one derived in Ref.~\cite{Morriss87}
which takes into account effects from inelastic collisions.

We next derive a generalized Jarzynski-type equality
\begin{equation}
\langle e^{\alpha \int_{0}^{t} ds \, \Omega(-s)} \rangle_{\beta} =
\langle e^{(\alpha-1) \int_{0}^{t} ds \, \Omega(s)} \rangle_{\beta}.
\label{eq:IFT-fractional}
\end{equation}
Since $\int_{0}^{t} ds \, \Omega(-s) = \int_{-t}^{0} ds \, \Omega(s)$,
the left-hand side can be expressed as
$\int d\tilde{\bf \Gamma} \, f_{\rm c}(\tilde{\bf \Gamma};\beta) \,
e^{\alpha \int_{-t}^{0} ds \, \Omega(\tilde{\bf \Gamma}(s))}$.
By setting $\tilde{\bf \Gamma} = {\bf \Gamma}(t)$, there holds,
since $\tilde{\bf \Gamma}(s) = {\bf \Gamma}(t+s)$,
\begin{equation}
\langle e^{\alpha \int_{0}^{t} ds \, \Omega(-s)} \rangle_{\beta} =
\int d{\bf \Gamma}(t) \, f_{\rm c}({\bf \Gamma}(t);\beta) \,
e^{\alpha \int_{0}^{t} ds \, \Omega({\bf \Gamma}(s))}.
\label{eq:IFT-fractional-dum-1}
\end{equation}
Using 
$H_{0}({\bf \Gamma}(t)) = 
H_{0}({\bf \Gamma}) + \int_{0}^{t} ds \, \dot{H}_{0}({\bf \Gamma}(s))$ and
$d{\bf \Gamma}(t) = e^{\int_{0}^{t} ds \, \Lambda({\bf \Gamma}(s))}
d{\bf \Gamma}$~\cite{Evans02}
which follows from the conservation of the number of ensemble members within a 
comoving phase volume, one derives
similarly to Eq.~(\ref{eq:new-neq-distribution})
\begin{equation}
d{\bf \Gamma}(t) \, f_{\rm c}({\bf \Gamma}(t);\beta) =
d{\bf \Gamma} \, f_{\rm c}({\bf \Gamma};\beta) \, e^{- \int_{0}^{t} ds \, \Omega({\bf \Gamma}(s))}.
\label{eq:IFT-fractional-dum-3}
\end{equation}
Substituting this into Eq.~(\ref{eq:IFT-fractional-dum-1}) yields
the equality~(\ref{eq:IFT-fractional}). 

For $\alpha = 1$, the equality~(\ref{eq:IFT-fractional})
reduces to the one noted below Eq.~(\ref{eq:Omega-def}).
By setting $\alpha = 0$ in Eq.~(\ref{eq:IFT-fractional}), one obtains
\begin{equation}
\langle e^{- \int_{0}^{t} ds \, \Omega(s)} \rangle_{\beta} = 1.
\label{eq:IFT}
\end{equation}
This equality is called the 
integral fluctuation theorem~\cite{Seifert05}
or the nonequilibrium partition identity~\cite{Evans08b},
originally derived as a corollary of the transient fluctuation theorem
that rests on microscopic time reversality.
Thus, the integral form of the fluctuation theorem holds also 
in the absence of such a symmetry.
Using the Jensen inequality, one obtains from Eq.~(\ref{eq:IFT})
\begin{equation}
\int_{0}^{t} ds \, \langle \Omega(s) \rangle_{\beta} \ge 0 
\,\,\, \mbox{for all} \,\,\, t
\label{eq:IFT-Jensen}
\end{equation}
which is referred to as the second-law inequality~\cite{Evans08b}.

The approach from an initial equilibrium towards a final steady
state can be characterized by the change in $\langle \Omega(t) \rangle_{\beta}$
from $\langle \Omega(0) \rangle_{\beta} = 0$ towards
$\langle \Omega \rangle_{\rm ss} = - \langle \Lambda \rangle_{\rm ss} \ge 0$
[see Eq.~(\ref{eq:SLLOD-Lambda})],
which follows from Eq.~(\ref{eq:Omega-def}) since
$\langle \dot{H}_{0} \rangle_{\rm ss} =
- \dot{\gamma} \langle \sigma_{xy} \rangle_{\rm ss} 
- 2 \langle {\cal R} \rangle_{\rm ss} = 0$
[see Eq.~(\ref{eq:dot-H0})].
Notice that the inequality~(\ref{eq:IFT-Jensen}) implies
$\langle \Omega(t) \rangle_{\beta} \ge 0$ for short $t$, 
meaning that the system must initially
move towards, rather than away from, the steady state.
This is an analogue of the statement originally made for
the relaxation to equilibrium
from nonequilibrium state~\cite{Evans08c}.

The fact that $\langle \Omega(t) \rangle_{\beta} \ge 0$ for short $t$
can also be derived from Eq.~(\ref{eq:dissipation-theorem-2})
specialized to $A = \Omega$, 
\begin{equation}
\langle \Omega(t) \rangle_{\beta} = \int_{0}^{t} ds \,
\langle \Omega(s) \Omega(0) \rangle_{\beta},
\label{eq:Omega-generalized-Green-Kubo}
\end{equation}
since the autocorrelation function must be positive at least for
short times.
If, in addition, the inequality $\langle \Omega(t) \Omega(0) \rangle_{\beta} \ge 0$ holds
for all $t$ as, e.g., in an exponential decay, 
one finds from Eq.~(\ref{eq:Omega-generalized-Green-Kubo}) that 
$\langle \Omega(t) \rangle_{\beta} \le \langle \Omega(t') \rangle_{\beta}$
for $t \le t'$, 
and the average dissipation function increases monotonically.
Since $\langle \Omega(t) \rangle_{\beta}$ is related to the
free energy production rate $\dot{\cal F}(t)$ via
$\beta \dot{\cal F}(t) = \langle \Omega(t) \rangle_{\beta}$~\cite{comment-free-energy-production},
such a monotonic increase in $\langle \Omega(t) \rangle_{\beta}$ implies that
the steady state can be characterized as the
state of maximum free energy production rate. 
However, the inequality $\langle \Omega(t) \Omega(0) \rangle_{\beta} \ge 0$ 
does not hold for all $t$ in general, and the
approach towards the steady state is not necessarily monotonic.
Again, this is an analogue of the statement for the relaxation to equilibrium~\cite{Evans08c}.

Finally,
we show that, at least formally, $f({\bf \Gamma},t)$
can be expressed in terms of the excess quantity. 
Such a representation has been derived exploiting the 
microscopic time reversality~\cite{Komatsu-all}, 
which however is broken here. 
Nevertheless, one can speak of paths 
evolving forward and backward in time
whose probabilities are related via the integral fluctuation theorem 
$\langle e^{\frac{1}{2} \int_{0}^{t} ds \, \Omega(-s)} \rangle_{\beta} =
\langle e^{-\frac{1}{2} \int_{0}^{t} ds \, \Omega(s)} \rangle_{\beta}$,
obtained by setting $\alpha = 1/2$ in Eq.~(\ref{eq:IFT-fractional}).
Utilizing such a symmetry, one can derive~\cite{Chong-PTP} 
\begin{equation}
f({\bf \Gamma},t)/f_{\rm c}({\bf \Gamma};\beta) =
\langle e^{\frac{1}{2} \int_{0}^{t} ds \, \Omega(-s)} \rangle_{{\bf \Gamma}; 0} /
\langle e^{- \frac{1}{2} \int_{0}^{t} ds \, \Omega(s)} \rangle_{{\bf \Gamma}; t} \, ,
\label{eq:KN-like-expression-0}
\end{equation}
where $\langle X \rangle_{{\bf \Gamma};t} \equiv
\int d\tilde{\bf \Gamma} X f_{\rm c}(\tilde{\bf \Gamma};\beta) 
\delta(\tilde{\bf \Gamma}(t) - {\bf \Gamma})/
f({\bf \Gamma},t)$
denotes the conditioned average.
We expand this expression
using cumulant average $\langle \cdots \rangle_{{\bf \Gamma};t}^{\rm c}$ to obtain
\begin{eqnarray}
& &
\log \frac{f({\bf \Gamma},t)}{f_{\rm c}({\bf \Gamma};\beta)} =
\frac{1}{2} 
\left\{
  \langle \Theta_{-}^{\rm ex} \rangle_{{\bf \Gamma}; 0} +
  \langle \Theta_{+}^{\rm ex} \rangle_{{\bf \Gamma}; t}
\right\} + 
\frac{t}{2} ( \bar{\Omega}_{-} + \bar{\Omega}_{+}) 
\nonumber \\
& & \quad 
+ \,
\sum_{k=2}^{\infty} \frac{1}{2^{k} k!}
\left\{
  \langle (\Theta_{-}^{\rm ex})^{k} \rangle_{{\bf \Gamma}; 0}^{\rm c} - (-1)^{k}
  \langle (\Theta_{+}^{\rm ex})^{k} \rangle_{{\bf \Gamma}; t}^{\rm c} 
\right\},
\label{eq:KN-like-expression-1}
\end{eqnarray}
in which
$\Theta_{\pm}^{\rm ex} \equiv
\int_{0}^{t} ds \, [ \Omega(\pm s) - \bar{\Omega}_{\pm}]$ with
$\bar{\Omega}_{\pm} \equiv \lim_{t \to \infty}
(1/t) \int_{0}^{t} ds \, \langle \Omega(\pm s) \rangle_{\beta}$.
Notice that $\langle \Theta_{\pm}^{\rm ex} \rangle_{\beta}/\beta$ is 
the excess free energy production (see above).

Following Ref.~\cite{Komatsu-all}, we
introduce a small parameter $\epsilon$ characterizing the 
``degree of nonequilibrium''.
Expressing all the quantities in natural dimensionless units,
we shall choose $\epsilon \sim \dot{\gamma} \sim \gamma$ so that 
$\Omega = O(\epsilon)$ 
[see Eqs.~(\ref{eq:sigma-def}), (\ref{eq:R-def}), (\ref{eq:SLLOD-Lambda}), and (\ref{eq:Omega-def})].
Applying Eq.~(\ref{eq:dissipation-theorem-2}) to 
$\langle \Omega(\pm t) \rangle_{\beta}$, one finds
$\bar{\Omega}_{-} + \bar{\Omega}_{+} = O(\epsilon^{3} t)$~\cite{Chong-PTP}.
The factor $\epsilon t$ comes from 
$e^{i {\cal L} t} = e^{i {\cal L}_{0} t} + O(\epsilon t)$,
where $i {\cal L}_{0}$ denotes the Liouville operator at $\epsilon = 0$.
Similarly, one finds that the last term in Eq.~(\ref{eq:KN-like-expression-1}) is
$O(\epsilon^{3} t)$.
Due to the additional $t$ dependence in the second term in 
Eq.~(\ref{eq:KN-like-expression-1}),
we obtain
\begin{equation}
\log \frac{f({\bf \Gamma},t)}{f_{\rm c}({\bf \Gamma};\beta)} =
\frac{1}{2} 
\left\{
  \langle \Theta_{-}^{\rm ex} \rangle_{{\bf \Gamma}; 0} +
  \langle \Theta_{+}^{\rm ex} \rangle_{{\bf \Gamma}; t}
\right\} + O(\epsilon^{3} t^{2}),
\label{eq:KN-like-expression-2}
\end{equation}
which resembles the representation derived in Ref.~\cite{Komatsu-all}.
The $t$ dependence is retained in the correction term since, e.g., in the non-Newtonian
regime where the relaxation time towards the steady state can become 
$\sim 1/\dot{\gamma} \sim 1/\epsilon$, 
the correction increases to $O(\epsilon)$ on that time scale. 
Thus, the expression~(\ref{eq:KN-like-expression-2}) can be useful only
if the steady state is reached on the time scale satisfying $\epsilon t \ll 1$.

However, such a naive small-$\epsilon$ expansion like Eq.~(\ref{eq:KN-like-expression-2})
might be questionable.
As noted below Eq.~(\ref{eq:dA-dbeta-dum-04}), one can choose an arbitrary initial
$\epsilon=0$ state for discussing the steady-state 
properties for $\epsilon > 0$.
Then, a question arises concerning what $\epsilon=0$ state the small-$\epsilon$
expansion should refer to.
We would like to leave the discussion on the $\epsilon=0$ versus $\epsilon \to 0$ state
to our future work. 

In this Letter, 
we derived the generalized Green-Kubo relation and the integral fluctuation theorem
for systems without microscopic time reversality and discussed their implications.
Our formulation for granular systems
will also be useful in studies of jammed glassy materials~\cite{jamming-all}.
In particular, the generalized Green-Kubo relation 
serves as a convenient exact starting point 
for approximate theoretical treatments,
and can easily be combined with liquid-state theories as detailed in Ref.~\cite{Chong09}. 

We thank S.~Sasa and H.~Tasaki for discussions. 
This work was supported by the Grant-in-Aid for scientific
research from the Ministry of Education, Culture, Sports,
Science and Technology (MEXT) of Japan 
(Nos.~20740245, 21015016, 21540384, and 21540388),
by the Global COE Program
``The Next Generation of Physics, Spun from Universality
and Emergence'' from MEXT of Japan,
and in part by the Yukawa International Program for
Quark-Hadron Sciences at Yukawa Institute for Theoretical
Physics, Kyoto University.
M.~O. thanks the Yukawa Memorial Foundation for financial
support.


\begin{thebibliography}{23}
\bibitem[{com({\natexlab{a}})}]{comment-mixing}
\bibinfo{note}{The system is said to exhibit mixing if arbitrarily chosen phase
  variables, say $A$ and $B$, become uncorrelated at long times, i.e., $\lim_{t
  \to \infty} \langle A(t) B(0) \rangle = \langle A(t) \rangle \langle B(0)
  \rangle$~\cite{Evans08}.}
\end{thebibliography}
\end{document}